\documentstyle[12pt,epsf]{article}
\begin{document}
\oddsidemargin1cm
\topmargin0cm
\headheight0cm
\headsep0cm
\topskip0cm
\textheight25cm
\textwidth17cm
\thispagestyle{empty}
\begin{flushright}
DO-TH 94/12 \\
May 1994
\end{flushright}
\vspace{2cm}
\begin{center}
  \begin{Large}
{\bf SEMILEPTONIC B-MESON DECAYS IN THE PARTON MODEL}
 \end{Large}
\end{center}
  \vspace{1cm}
   \begin{center}
C.\ H.\ Jin$^a$\footnote{Supported by Bundesministerium f\"ur Forschung
und Technologie, 05-6DO93P, Bonn, FRG and by the CEC Science Project
$n^{o}$ SC1-CT91-0729.},
W.\ F.\ Palmer$^b$, E.\ A.\ Paschos$^a$$^1$\\
      \vspace{0.3cm}
        $^a$Institut f\"ur Physik\\  Universit\"at Dortmund\\
        D--44221 Dortmund, Germany\\
  \vspace{0.3cm}
        $^b$Department of Physics, The Ohio State University \\
        Columbus
, Ohio 43210, USA\\
\end{center}
\vspace{2.0cm}
\begin{abstract}
We study the inclusive semileptonic B decays in the parton model. The
model gives a decay spectrum which is in excellent agreement with
experiment and makes several predictions to be tested in the forseeable
future.
\end{abstract}
\vfill
\begin{center}
Contribution to the XXIXth Rencontres de Moriond,\\
  {\it Electroweak Interactions and Unified Theories}, \\
M\'eribel, France, March 12--19, 1994.
\end{center}
\newpage
\renewcommand{\baselinestretch}{1.5}
\normalsize
Semileptonic B-meson decays have been studied in our group for a long time
[1-3] and
I am happy to present our latest results at this Moriond meeting.
There are two reasons
for presenting the results here: (1) We have several new predictions which can
be tested experimentally, and (2) There are experimental colleagues at this
meeting, who may be interested in testing the model. For
the foreseeable future B-meson
physics will be a popular field of research with various experiments planned
at the
following accelerators: CESR, LEP, Tevatron, HERA-B and the two new
B-factories: at SLAC (USA) and at KEK (Japan).

The oldest and most popular model for analysis of semileptonic B-decays
is the ACCMM model [4]. This model improves the
free quark decay by taking into account the Fermi motion of the spectator
quark in the B-meson. Related work along this line of research
analyses the decay with bound state
wave functions [5].

The second model is the parton model developed in direct analogy to deep
inelastic scattering with a short distance contribution described by
the singularities of products of operators on the light-cone and
a long distance part described by the
distribution function for the b-quark in the B-meson. We have recently
established that the model is in excellent agreement with the
data and it makes many predictions to be tested in future experiments [3].

A third approach is the heavy quark effective theory (HQET) which hopes to
calculate the decay from first principles [6]. The method, so far, has met
with partial success because of technical difficulties.
A success of HQET is the calculation of
corrections for the interaction of the decay products, which include
non-perturbative effects and turn out to be numerically small.
On the other hand, the lepton spectra derived by this method are singular
at the edge of the phase space. Thus modelling the end-point region of
the phase space is still necessary for eliminating the singularities and
for the extraction of $V_{ub}$, as emphasized by Manohar and Wise in
reference [6].
The parton model is one of several approaches that can describe this
region [1-3].  Others have been developed in references
[7,8].

In the parton model we visualize the decay to proceed at an infinite momentum
frame with the
decay width being the decay of a free quark times the probability $f(x)$
of finding b-quark carrying momentum $xP_B$ within the B-meson.
The distribution function $f(x)$ is
 not Lorentz
invariant and is defined at the infinite momentum frame. It can not be
transformed to the rest frame of the B-meson because it involves an
infinite sum of quark-antiquark pairs whose calculation requires a
complete solution of the field theory.
For our analysis we need the
kinematic variables:  $P_{B}$=  momentum of B-meson;
$P_{e}, E_e$    		=    momentum, energy of electron;
$P_{\nu}$               =    momentum of neutrino;
$P_{X}$                 =    momentum of hadrons;
$q=P_{e}+P_{\nu}$       =    momentum of current; and
$M_X$                   = invariant mass of the
final hadronic system.
Here we give explicit formulas for $b\rightarrow u$ decays and those for
$b\rightarrow c$ decays can be obtained using the similar method [9].

We follow the standard analysis of writing the
decay probability as the Fourier transform of the current commutator. The
lowest dimension operator gives the leading singularity on the
light-cone, which is
\begin{equation}
[j_{\mu}(y), j_{\nu}^{\dagger}(0)]=\frac{1}{\pi}(-S_{\mu\alpha\nu\beta}+
i\varepsilon_{\mu\alpha\nu\beta})j^{\beta}(0,y)\partial^\alpha
[\varepsilon(y_{0})\delta(y^2)].
\end{equation}
Its Fourier transform leads to a $\delta$-function
\begin{equation}
\varepsilon(xP_{B0}-q_{0})\delta(x^{2}P_{B}^{2}-2xq\cdot P_{B}+q^{2})=
\frac{1}{2M_{B}\left| \vec q \right|}[\delta(x-x_{+})-\delta(x-x_{-})],
\end{equation}
showing two scaling variables
\begin{equation}
x_{\pm}=\frac{q_{0}\pm \left| \vec q \right|}{M_{B}}=\frac{q_{\pm}}{M_{B}} ,
\end{equation}
which are the light-cone variables of the current-momentum. The second
root appears for the first time in the decays of heavy particles and is a
 consequence of field
theory. It corresponds to the creation of a quark-antiquark pair in the
B-meson through Z-diagram.
We define the distribution function $f(\xi)$ through the Fourier transform
of the matrix element of the bilocal current
\begin{equation}
\langle B\left|j^{\beta}(0, y)\right| B\rangle= 4\pi P_B^{\beta}F(y\cdot P_B)
= 2P_B^{\beta}\int_{-\infty}^{\infty}d\xi
e^{-i\xi y\cdot P_B}f(\xi) .
\end{equation}

We represent the decay with the diagrams in Figs.~1 and 2, which look
very much like those in deep inelastic scattering. We see in Fig.~2 that
the non-perturbative effects are contained in the elliptical regions A, B
and C.
The effects from A and B are included in the distribution function.
Region C includes final state interactions of the emerging light quark.
It can be viewed as the interaction of the light quark with the
background gluonic field [6].
\vspace{4.5cm}
\begin{center}
Fig.~1 Semileptonic B-decays in the parton model.
\end{center}
\vspace{3cm}
\begin{center}
Fig.~2 The two-current correlator.
\end{center}

Using standard methods we obtained the triple differential
decay rate of $b\rightarrow ue\bar \nu$
\begin{equation}
\frac{d\Gamma}{dE_{e}dydx_{+}}= \frac{G_{F}^{2}M_{B}^{4}
\left| V_{ub}\right|^{2}}
{8\pi^{3}}y\lbrace
x_{+}f(x_{+})(x_{+}-y) - (x_{+} \leftrightarrow x_{-})\rbrace ,
\end{equation}
where $y=2E_{\nu}/M_{B}$.
The second term with $x_{-}$ will be
neglected because it can be shown that for both $b\rightarrow u$ and
$b\rightarrow c$ decays $x_{-} \leq 1/2$ and the distribution
function has a peak in the region $0.7< x <1$.
We see a remarkable factorization property that the distribution functions
depend primarily on $x_{+}$ and the $y$
dependence is explicitly shown.  A similar formula was
derived for the $b\rightarrow ce\bar \nu$ decays by keeping the mass of
the charm quark [9].

The above formula is analogous to the formula in deep inelastic
neutrino-nucleon scattering and we comment on its properties.

(I) The fact that the structure function for $B\rightarrow X_{u}l\bar \nu$
is actually a function of a single dimensionless variable
$x_{+}=(q_{0}+q_{z})/M_B$ is a very strong statement analogous to Bjorken
scaling in deep inelastic scattering. This scaling behavior in the
inclusive decays of B mesons was emphasized in several papers [1-3].
Recently, a scaling was also suggested in HQET [8].

(II) The dependence on the variables $E_e$ and $y$ is explicit in eq.~(5).
For fixed $x_+$ the $y$ distribution is a parabola with the maximum at
$y= x_{+}/2$. The family of the y-distributions is shown in Fig.~3.
This property can be used to measure the distribution function $f(x)$
experimentally.
\epsfysize=6cm
\centerline{\epsffile{triple.eps}}
\begin{center}
Fig.~3 The y-distribution in B-decay for various values of $x_+$.
\end{center}

(III) When we integrate eq.~(5) over $x_+$ we obtain
\begin{equation}
\frac{d\Gamma}{dE_edy}= \frac{G_F^2M_B^4\left| V_{ub}\right|^2}{8\pi^3}
[\langle x_+^2\rangle y- \langle x_+\rangle y^2]
\end{equation}
for $b\rightarrow u$ decays and a similar formula for $b\rightarrow c$.
We see now the explicit dependence on $y$ weighted by the moments
$\langle x_+\rangle$ and $\langle x_+^2\rangle$. The moments may depend
on $y$ through the limits of integration and require some attention. We
note that in eqs.~(5) and (6) there is only the mass of the B meson.
There is no mass of the b quark because of the standard substitution
$p_b=xP_B$.

(IV) We calculated the electron spectrum for the decay $B\rightarrow
X_ce\bar \nu$ and compared it with the published measurements of the
ARGUS group [10]. In Fig.~4 we show the data and our calculation to
be explained below. The agreement between theory and experiment is
impressive. This encouraged us to develop additional tests for the theory.

(V) As high statistics data becomes available we can probe the decay dynamics
in a deeper way by studying double differential decay distributions.
At the end of the article we present detailed distributions.

The properties (I)-(III) presented above are general and independent of
the distribution function. For the results in (IV) and (V) we need the
distribution function as a function of the scaling variable $x_+$.
There are physical arguments relating the distribution function to the
fragmentation function
\vspace{0.2cm}
\epsfysize=7cm
\centerline{\epsffile{compare2.eps}}
\begin{center}
Fig.~4 The electron spectrum and its comparison with the ARGUS data.
\end{center}
of a b-quark producing a B-meson. We shall adopt
this identification and use the form
\begin{equation}
f(x)= N\frac{x(1-x)^{2}}{\lbrack (1-x)^{2}+\varepsilon_{p}x\rbrack^{2}}
\end{equation}
with $\varepsilon_{p}$ a parameter in the range $0.003-0.009$
and $N$ the normalization factor. The function $f(x)$ peaks at large
values of $x$ as expected for a meson composed of a heavy quark and a light
quark. With this distribution function we obtain the spectrum shown in
Fig.~4. The parameters that we used
 are $\left|V_{cb}\right|= 0.042$,
$\varepsilon_p=0.003$, $m_c= 1.4 \ GeV$ with the corresponding
normalization factor $N= 0.088$.
We have also calculated the total semileptonic width and found that it
is by $15\%$ smaller than the prediction of the free quark model.
This reduction comes from the moments of $x_+$ which appear in eqs.~(5)
and (6).
A comparison with the HQET spectrum has
not being made yet because of the presence of singularities at the edge
of the phase space. The ACCMM model gives also
a good fit for $B\rightarrow X_ce\bar \nu$ decays. We compared the two
models by computing the $b\rightarrow u$ decays for the same parameters
and the results are shown in Fig.~5. We notice that the parton
model
spectrum is softer than the ACCMM model spectrum.

For double differential decay distributions one will be able to measure
the total hadronic energy of the final state $E_X= M_B-q_0$ in addition
to the electron energy. This will be easiest at the asymmetric B-factories
where the two hadronic jets are spatially separated.
In Fig.~6
we show the $b \rightarrow u$
double differential decay rate in $E_{e}$ and $q_{0}$. The
spectra show a striking dependence of  $q_{0}$. Most of the events occur for
$0.3 M_{B}\leq E_{X}\leq 0.5 M_{B}$. For smaller hadronic energies the
spectra shift to higher electron energies, $E_{e}\geq 2.0 \ GeV$. This
correlation of events may help to isolate $b\rightarrow u$ events.

\epsfysize=5.7cm
\centerline{\epsffile{QCDE.eps}}
Fig.~5 A comparison of the electron spectra for $b\rightarrow u$ decays
in ACCMM model and parton model.

\epsfysize=6.1cm
\centerline{\epsffile{eq.eps}}
\begin{center}
Fig.~6 The double differential decay rate for $b\rightarrow u$ decays.
\end{center}

For comparison we also calculated the distribution for the
$B\rightarrow X_{c}+e^{-}\bar\nu$ decay. In Fig.~7 we show the double
differential decay rate in $E_{e}$ and $q_{0}$. Comparing with Fig.~6 we
note that this channel runs out of events at  $E_{e}=2.3$ $GeV$ and the events
above 2 $GeV$ are very few. By making a cut in $q_{0}>M_{B}-M_{D}=0.65M_{B}$
there are no events left for $b\rightarrow c$. This can be used as another
criterion for isolating $b\rightarrow u$ events.

The parton model provides a powerful method for analysing inclusive
semileptonic B decays. It gives a complete describtion of the decay
in several kinematic variables and makes predictions accessible to
experiments. It will be very interesting to compare detailed spectra
with the predictions of the model.

\epsfysize=6.1cm
\centerline{\epsffile{eq2.eps}}
\begin{center}
Fig.~7 The same as Fig.~6 but for $b\rightarrow c$ decays.
\end{center}
\renewcommand{\baselinestretch}{1}
{\bf References}\newline
[1] A. Bareiss and E.A. Paschos, {\it Nucl. Phys.}
{\bf B327} (1989) 353\newline
[2] A. Bareiss, {\it Z. Phys.} {\bf C53} (1992) 311 \newline
[3] C.H. Jin, W.F. Palmer, and E.A. Paschos,
preprint DO-TH 93/21 and OHSTPY-HEP-T-93-011, 1993 (unpublished);
C.H. Jin, W.F. Palmer, and E.A. Paschos, to appear in
{\it Phys. Lett.} {\bf B} \newline
[4] G. Altarelli, N. Cabibbo, G. Corbo, L. Maiani, and
G. Martinelli, {\it Nucl. Phys.} {\bf B208} (1982) 365 \newline
[5] A.V. Dobrovolskaya, K.A. Ter-Martirosyan, and A.I. Veselov,
{\it Phys. Lett.} {\bf B257} (1991) 490\newline
[6] J. Chay, H. Georgi, and B. Grinstein, {\it Phys. Lett.}
{\bf B247} (1990) 399;
 I.I. Bigi, M.A. Shifman,
N.G. Uraltsev, and A.I. Vainshtein, {\it Phys. Rev. Lett.} {\bf 71}
(1993) 496;
 A.V. Manohar and M.B. Wise, {\it Phys. Rev. }
{\bf D49} (1994) 1310;
T. Mannel, {\it Nucl. Phys.} {\bf B413} (1994) 396 , for more references
see [3]\newline
[7] M. Neubert, preprint CERN-TH.7087/93\newline
[8] I.I. Bigi, M.A. Shifman, N.G. Uraltsev, and A.I. Vainshtein,
preprint CERN-TH.7129/93 (1993)\newline
[9] C.H. Jin, submitted to the XXVII International
Conference on High Energy Physics, Glasgow, 1994\newline
[10] H. Albrecht {\it et al.} (ARGUS), {\it Phys. Lett.}
{\bf B318} (1993) 397
\end{document}